\documentclass[10pt,draft]{dis03}
\usepackage{epsf,amsmath}

\newcommand{\bgi}{\begin{itemize}}
\newcommand{\eni}{\end{itemize}}

\newcommand{\bb}{\begin{thebibliography}}
\newcommand{\eb}{\end{thebibliography}}

\newcommand{\nn}{\nonumber \\}
\newcommand{\bwt}{\begin{widetext}}
\newcommand{\ewt}{\end{widetext}}
\newcommand{\bea}{\begin{eqnarray}}
\newcommand{\ba}{\begin{array}}
\newcommand{\ea}{\end{array}}
\newcommand{\eea}{\end{eqnarray}}
\newcommand{\be}{\begin{equation}}
\newcommand{\ee}{\end{equation}}

\newcommand{\bit}[1]{\bibitem{#1}}

\newcommand{\nnn}{\nonumber}

\newfont{\fib}{cmfi10 at 10pt}

\begin{document}

\title{Nucleon Transversity Properties Through $ep$ Scattering\thanks{On 
the basis of the talks delivered by K.A.~Oganessyan (kogan@mail.desy.de) 
at DIS2003, XI International Workshop on Deep Inelastic Scattering, 
St. Petersburg, 23-27 April, 2003.}}

\author{
L.~P.~Gamberg \\
Division of Science, Penn State-Berks Lehigh Valley College, \\ 
Reading, PA 19610, USA \\
G.~R.~Goldstein\\ 
Department of Physics and Astronomy, Tufts University,\\
Medford, MA 02155, USA  \\
K.~A.~Oganessyan\\
INFN-Laboratori Nazionali di Frascati, E. Fermi 40, \\ 
00044 Frascati, Italy \\
DESY, Deutsches Elektronen Synchrotron Notkestrasse 85, \\ 
22603 Hamburg, Germany }

\maketitle

\begin{abstract}
\noindent The $T$-odd distribution functions contributing to transversity
properties of the nucleon  and their role in fueling nontrivial contributions 
to azimuthal asymmetries in  semi-inclusive deep inelastic scattering
are investigated. 
We use a dynamical model to evaluate these quantities in
terms of HERMES kinematics. We point out how the measurements of 
$\cos2\phi$ asymmetry may indicate the presence of 
$T$-odd structures in unpolarized $ep$ scattering.
\end{abstract}

\section{Introduction} 

It is widely recognized that the distributions in the 
azimuthal angle $\phi$ of the detected hadron in hard scattering 
processes provide interesting variables to study in both perturbative 
and non-perturbative regimes. They are of great interest since they test 
perturbative quantum chromodynamics (QCD) predictions for the 
short-distance part of strong interactions and yield important 
information on the long-distance internal structure 
of hadrons computed in QCD by non-perturbative methods. 
These nonperturbative contributions  are parameterized 
effectively by introducing transverse (so-called 
``intrinsic'' transverse motion) degrees of freedom in the parton 
distribution and fragmentation functions. The different combinations of 
the transverse momentum and spin results{\bf ,} in a rich variety of 
information on the hadron spin structure. In particular, time 
reversal odd ($T$-odd) structures~\cite{siv90,col93,cplb,tm95,kotz96,bm98} 
can appear -- they
filter the novel transversity properties of quarks in hadrons. Such 
structures are accessible through the  azimuthal asymmetries in semi-inclusive
and polarized spin processes. 
Beyond the $T$-odd properties, the existence of these distributions
is a signal of the {\em essential} role 
played by the intrinsic transverse quark
momentum and the corresponding angular momentum of quarks inside
the target and fragmenting hadrons  in these hard scattering processes.  
In this paper we analyze the $T$-odd transversity properties 
of quarks
in hadrons which emerge in semi-inclusive deep inelastic scattering (SIDIS). 
We apply these results to 
predict $\cos2\phi$~\cite{bm98} and $\sin(\phi-\phi_S)$ Sivers~\cite{siv90} 
asymmetries 
in terms of HERMES kinematics. Also, we discuss how the measurements of 
$\cos2\phi$ asymmetry may indicate the existence of 
$T$-odd structures in spin-independent semi-inclusive $ep$ scattering.

\section{T-Odd Distributions in Semi-Inclusive Reactions }

Recently  rescattering was considered as a mechanism 
for SSAs in pion electroproduction from transversely polarized
nucleons.  Using the QCD motivated quark-diquark model of the 
nucleon~\cite{rodriq,bhs}, the $T$-odd distribution function,
$f_{1T}^\perp(x,k_\perp)$ (representing the number density of 
{\em unpolarized} quarks in transversely polarized nucleons) and  
the corresponding analyzing power for 
the azimuthal asymmetry
in the fragmenting hadron's momentum and spin distributions
resulted in a leading twist nonzero Sivers 
asymmetry~\cite{ji,cplb}.  Using the approach in Ref.~\cite{ji},
we  investigated the 
rescattering in terms of initial/final state interactions  
to the $T$-odd function $h_1^\perp(x)$ 
and corresponding azimuthal asymmetry in 
SIDIS~\cite{gold_gamb,gamb_gold_ogan,gamb_gold_ogan1}. 
The  asymmetry involves the convolution with the $T$-odd 
fragmentation function, $h_1^\perp(x) \otimes H_1^\perp(z)$~\cite{bm98}. 
The function 
$h_1^\perp(x,k_\perp)$ (representing 
the number density of {\em transversely polarized} quarks in unpolarized
nucleons) is complimentary to the Sivers function and
is of great interest theoretically,  since it vanishes at tree 
level, and experimentally, since its determination does not 
involve polarized nucleons~\cite{bm98,cplb,gold_gamb,gamb_gold_ogan,
gamb_gold_ogan1,bbh}.  

The $T$-odd distributions are readily defined from the 
transverse momentum dependent quark distributions~\cite{col82,cplb} 
where the well known identities for manipulating the limits of
an ordered exponential lead to the expression
\bea
\Phi^{[\Gamma]}(x,k_\perp)&=&{\frac{1} {2}}\sum_n
\int {\frac{d\xi^- d^2\xi_\perp }
  {(2\pi)^3}} e^{-i(\xi^- k^+-\vec{\xi}_\perp \vec{k}_\perp)} 
\nn  && \hspace{-2.7cm}
\times
\langle P|\overline{\psi}(\xi^-,\xi_\perp){\cal G}^{\dagger}(\infty,\xi)
\big|n\rangle\Gamma\langle n\big|
{\cal G}(\infty,0)\psi(0)|P\rangle\vert_{{\scriptscriptstyle{\xi^+=0}}}
\label{link}
\eea
and the path ordered exponential is 
\bea
{\cal G}(\infty,\xi)={\cal P}
\exp{\left(-ig\int_{\xi^-}^\infty d\xi^- A^+(\xi)\right)},
\nnn
\eea
and $\{\big|n\rangle \}$ are a complete set of states.
While the path ordered light-cone link operator is necessary to maintain 
gauge invariance and appears to respect factorization~\cite{cplb,ji,ji2} 
when transverse momentum distributions are considered, 
in non-singular gauges~\cite{ji,ji2}, it also provides a mechanism
to generate  interactions between an eikonalized struck
quark and the remaining target.
These final state interactions in turn give rise to leading 
twist contributions to the distribution functions that fuel the 
novel SSAs that have been reported in the 
literature~\cite{bhs,cplb,ji,ji2,gold_gamb,bbh,gamb_gold_ogan,
gamb_gold_ogan1}. 
It is worth  mentioning that the $T$-odd distribution functions in 
SIDIS and Drell-Yan are not equal and have opposite signs~\cite{cplb,
metz}. New experimental results will test the issue of universality of 
distribution functions.

The quark-nucleon-spectator model used in previous rescattering 
calculations assumed a point-like nucleon-quark-diquark vertex, which 
leads to logarithmically divergent, $x$-dependent distributions.   
To address the $\log$ 
divergence~\cite{bhs,ji,gold_gamb,bbh,gamb_gold_ogan} we
assume the transverse momentum dependence of the quark-nucleon-spectator 
vertex can be approximated by a Gaussian distribution in 
$k_\perp^2$~\cite{gamb_gold_ogan1}. 
Performing the loop integration,
and projecting the unpolarized
piece from $\Phi^{\scriptscriptstyle [i\sigma^{\perp +}\gamma_5]}$    
results in the
leading twist, $T$-odd, unpolarized contribution
\bea
\Phi^{\scriptscriptstyle [\sigma^{\perp +}\gamma_5]}_{\scriptscriptstyle 
[h_1^\perp]}\hspace{-.15cm}&=&\hspace{-.15cm} 
\frac{\varepsilon_{
\scriptscriptstyle 
+-\perp j}k_{\scriptscriptstyle \perp j}}{M}h_1^\perp(x,k_\perp)
\nn &=&\frac{e_1e_2g^2}{2(2\pi)^4}\frac{b^2}{\pi^2}
\frac{(m+xM)(1-x)}{\Lambda(k^2_\perp)}
\frac{\varepsilon_{\scriptscriptstyle +-\perp j}
k_{\scriptscriptstyle \perp j}}{k_\perp^2}
\nn &&\times
e^{-b\left(k^2_\perp- \Lambda(0)\right)}
\hspace{-.15cm}\left[\Gamma(0,b\Lambda(0))\hspace{-.10cm}-\hspace{-.10cm}
\Gamma(0,b\Lambda(k^2_\perp))\right] .
\nn
\eea
Here, $e_1$ ($e_2$) is the charge of the struck quark (gluon-scalar diquark 
coupling), and 
$\Lambda(k^2_\perp)=k_\perp^2 +(1-x)m^2 +x\lambda^2  -x(1-x)M^2$, where 
$M$, $m$, and $\lambda$ are the nucleon, quark, and diquark masses 
respectively.  Also, $b=\frac{1}{<k_\perp^2>}$, where  $<k_\perp^2>$ 
is fixed below.    
The average $k^2_{\perp}$ is a regulating scale 
which we fit to the expression for the integrated unpolarized 
structure function
\bea
f(x)&=&\frac{g^2}{(2\pi)^2}\frac{b^2}{\pi^2}\left(1-x\right)
\nn &&
\times\Bigg\{\frac{\left(m+xM\right)^2-\Lambda(0)}{\Lambda(0)}
\nn &&\hspace{-1cm}
-\left[2b\left(\left(m+xM\right)^2-\Lambda(0)\right)-1\right]
e^{2b\Lambda(0)}\Gamma(0,2b\Lambda(0))\Bigg\}\, ,
\nn
\eea
which multiplied by $x$ at $<k_\perp^2> = {(0.4)}^2$ GeV$^2$ 
is in  good agreement with the valence
distribution of Ref.~\cite{GRV}.
 
\section{Azimuthal Asymmetries}

We discuss the explicit results and numerical evaluation of the
spin-independent double $T$-odd $\cos 2\phi$ and single 
transverse-spin $\sin (\phi-\phi_s)$ asymmetries for
$\pi^+$ production in SIDIS.
\begin{figure}[!thb]
\begin{center}
\centerline{\epsfxsize=2.3in\epsfbox{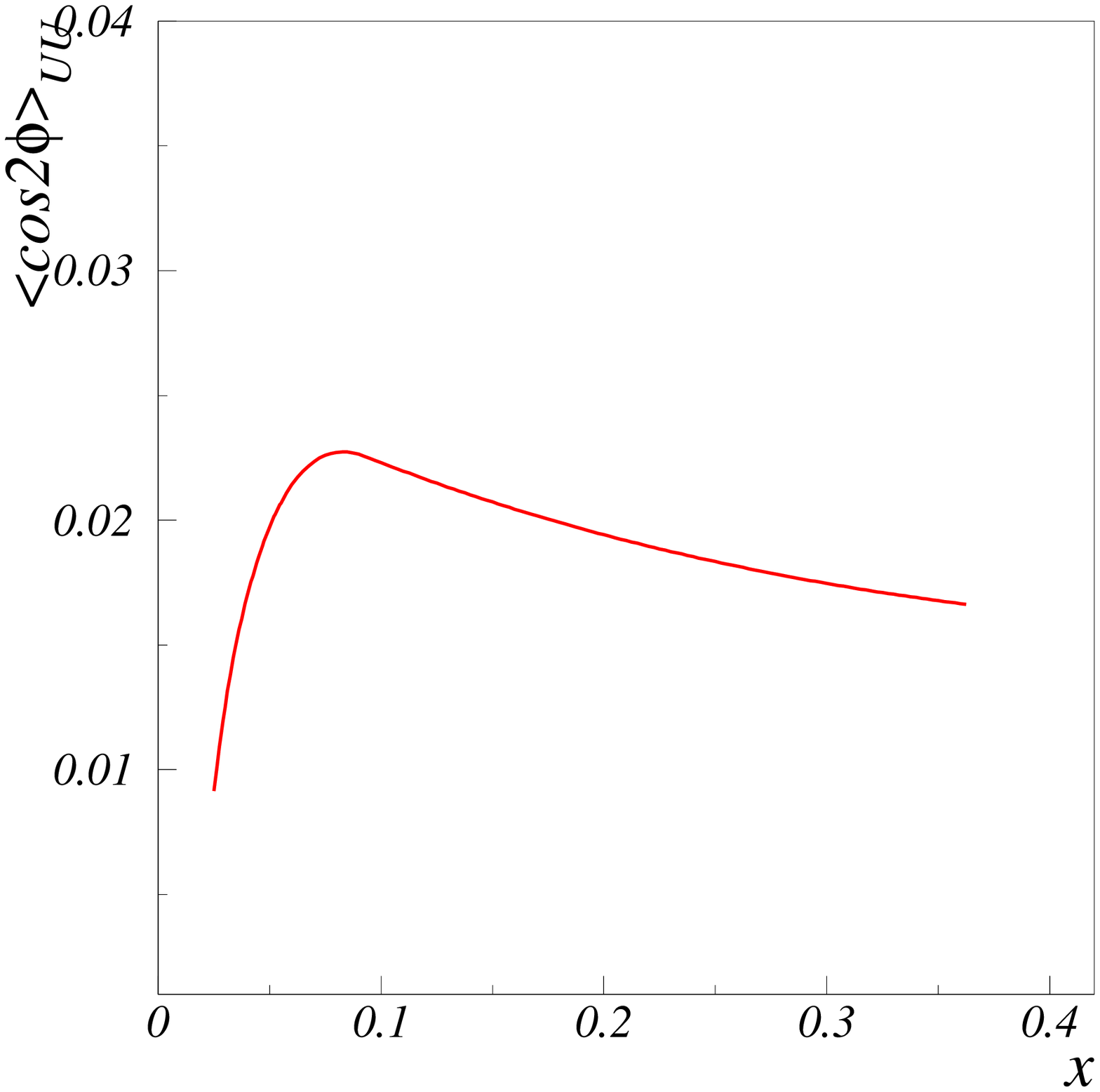}
\epsfxsize=2.3in\epsfbox{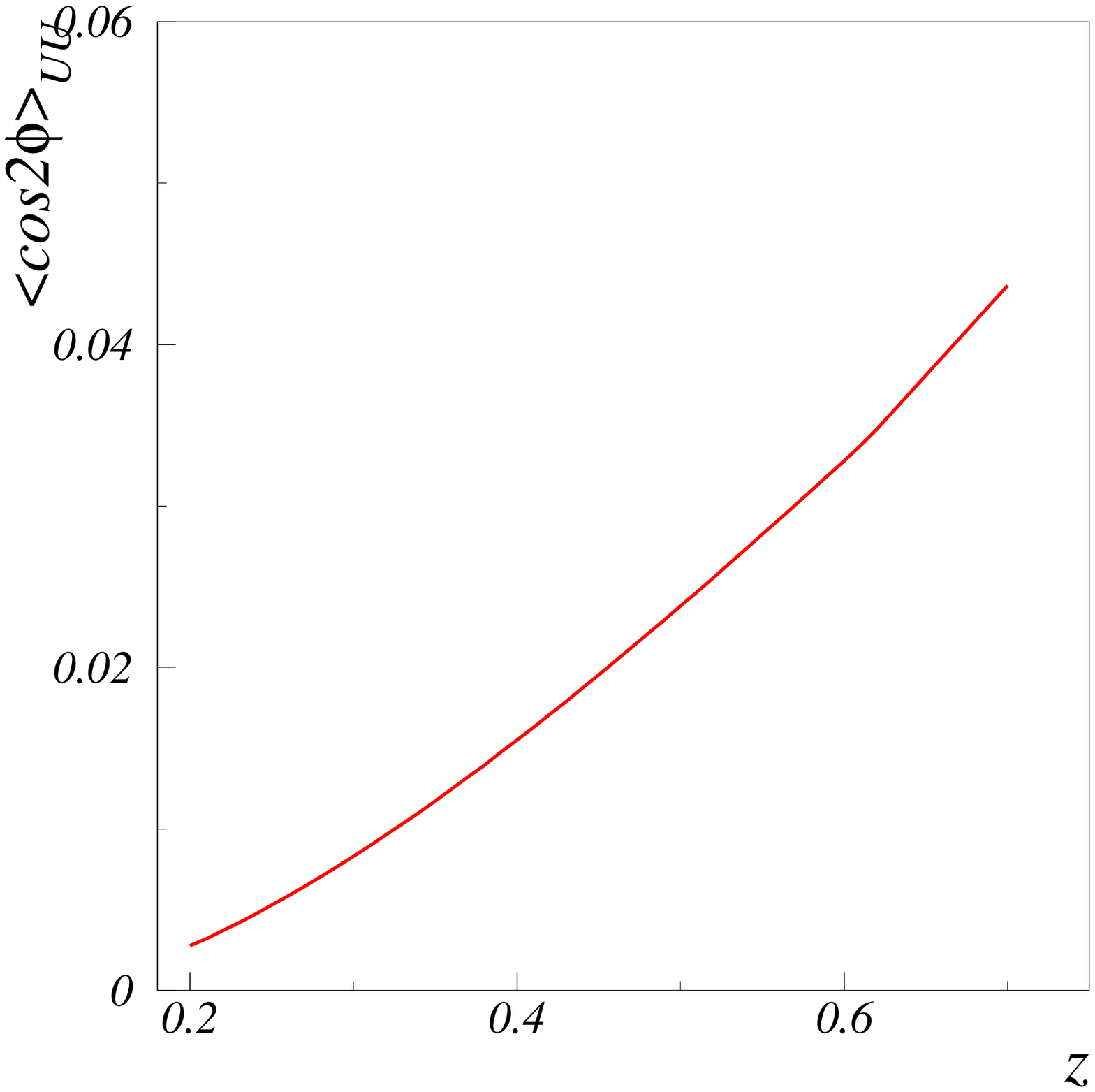}}
\caption[*]{ \label{A}Up Panel: The \protect{${\langle \cos2\phi 
\rangle}_{\scriptscriptstyle UU}$} asymmetry for \protect{$\pi^+$} 
production as a function of \protect{$x$}. Down Panel: The 
\protect{${\langle \cos2\phi \rangle}_{\scriptscriptstyle UU}$} asymmetry for 
\protect{$\pi^+$} production 
as a function of \protect{$z$}.}
\end{center}
\end{figure}
We use the conventions established in~\cite{bm98} for the
asymmetries.  Being $T$-odd, $h_1^\perp$ appears with the $H_1^\perp$, 
the $T$-odd fragmentation function in observable quantities.
In particular, the following weighted 
SIDIS cross section projects out a leading double $T$-odd
$\cos2\phi$ asymmetry, 
\bea
{\langle \frac{\vert P^2_{h{\perp}} \vert}{M M_h} \cos2\phi \rangle}
{\scriptscriptstyle_{UU}}&=& 
\frac{\int d^2P_{h\perp} 
\frac{\vert P^2_{h\perp}\vert}{M M_h}
\cos 2\phi\,  d\sigma}
{\int d^2 P_{h\perp}\, d\sigma} 
\nn &=&\frac{{8(1-y)} \sum_q e^2_q h^{\perp(1)}_1(x) z^2 H^{\perp(1)}_1(z)}
{{(1+{(1-y)}^2)}  \sum_q e^2_q f_1(x) D_1(z)}
\label{ASY} 
\nn
\eea
where the subscript $UU$ indicates unpolarized beam and 
target. The non-vanishing 
$\cos2\phi$ asymmetry originating from $T$-even distribution and 
fragmentation function 
appears at order $1/Q^2$~\cite{CAHN,kotz96,OABD}. This consideration 
will be discussed  in the next section. 

\begin{figure}[!thb]
\begin{center}
\centerline{\epsfxsize=2.3in\epsfbox{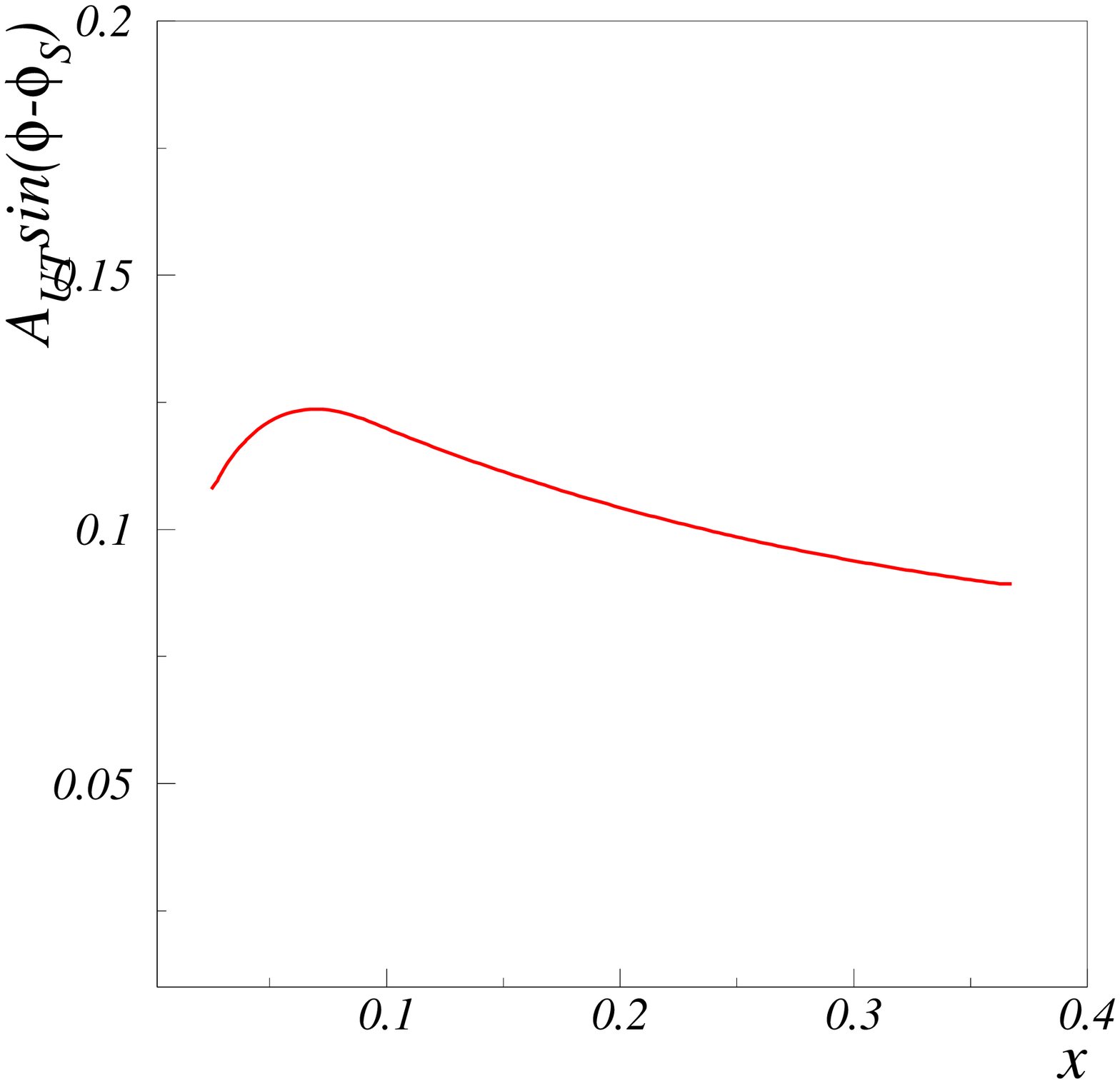}
\epsfxsize=2.3in\epsfbox{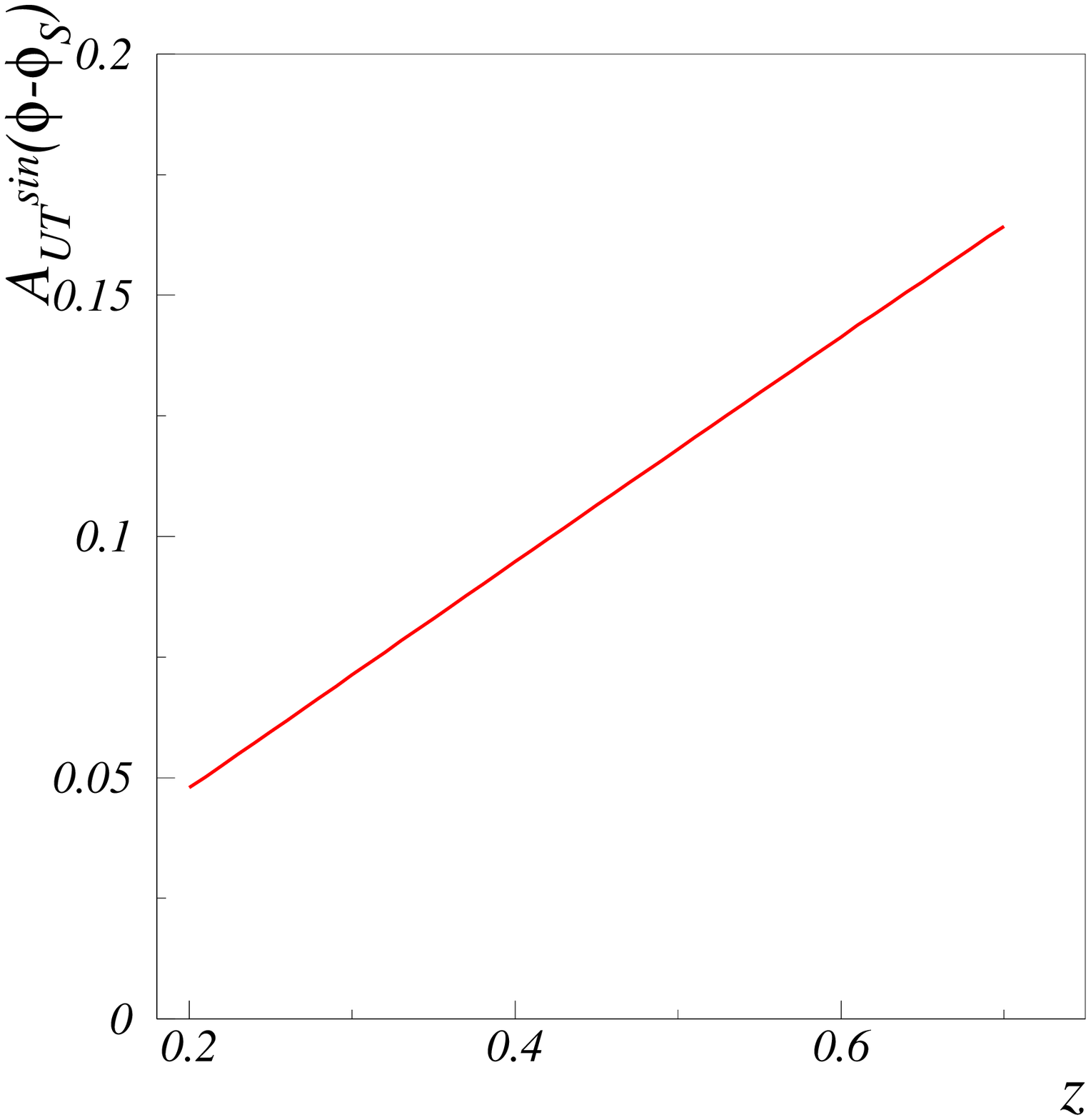}}
\caption[*]{\label{sivers} Up Panel: The 
\protect{$A^{\scriptscriptstyle\sin(\phi-\phi_S)}_{\scriptscriptstyle UT}$ 
$x$ }
dependent Sivers asymmetry.
Down Panel: The 
\protect{$A^{\scriptscriptstyle\sin(\phi-\phi_S)}_{\scriptscriptstyle UT}$
 $z$} dependent Sivers asymmetry.}
\end{center}
\end{figure}

Additionally, the SSA characterizing the
so-called Sivers effect is
\bea
\langle \frac{\vert P_{h\perp} \vert}{M} 
\sin(\phi-\phi_S) \rangle_{\scriptscriptstyle UT}
&=& 
\frac{\int d^2P_{h\perp} \frac{\vert P_{h\perp}\vert}{M}
\sin (\phi-\phi_S) \, d\sigma}
{\int d^2 P_{h\perp} \, d\sigma} 
 \nn  &&  \hspace{-2cm} = 
\frac{(1+{(1-y)}^2) \sum_q e^2_q f^{\perp(1)}_{1T}(x) z D^q_1(z)}
{{(1+{(1-y)}^2)}  \sum_q e^2_q f_1(x) D_1(z)},
\label{ASY1} 
\nn
\eea
where the subscript $UT$ indicates unpolarized beam and transversely
polarized target.
The functions $h_1^{\perp (1)}(x)$, $f_{1T}^{\perp (1)}(x)$, and 
$H_1^{\perp (1)}(z)$ are the weighted
moments of the distribution and fragmentation functions~\cite{km97}.  

In Figs.~\ref{A} and ~\ref{sivers} the results from  
Ref.~\cite{gamb_gold_ogan1} for the ${\langle \cos2\phi 
\rangle}_{UU}$ and $A^{\sin(\phi-\phi_S)}_{UT}$ for $\pi^+$ production 
on a proton target, evaluated for HERMES kinematics, are presented as a
function of $x$ and $z$, respectively. 

\section{The phenomenology of $\cos2\phi$ asymmetry}

The effects that vanish as $M^2/Q^2$ 
are important at  small and moderate values of $Q^2$. Such 
effects can arise in the Feynman-Bjorken model. The best known 
of these is the ratio of the longitudinal to transverse cross 
section~\cite{fmn72}:
\begin{equation}
R=\frac{\sigma_S}{\sigma_T}=\frac{4(m^2+\langle k^2_{\perp} \pm 
\Delta^2 \rangle)}{Q^2},
\label{FB}
\end{equation}
where $m$ is the parton mass, and $\Delta$ an unknown 
correction due to possible parton-parton interactions. The ratio is zero 
in the standard limit of $Q^2 \to \infty$, but cannot be ignored for 
moderate $Q^2$. 

The differential cross section for the process $ep \to e'hX$ involves four 
structure functions~\cite{cheng72}:
\begin{equation}
d\sigma \propto W^{++}+(1-y)W^{00}
+(1-y)W^{+-}\cos2\phi+\sqrt{(1-y)(2-y)}Re(W^{+0})\cos\phi,
\label{CS}
\end{equation}
where $+,-,0$ correspond to photons with positive, negative, and zero 
helicity. 

\begin{figure}[!thb]
\centerline{\epsfxsize=3.0in\epsfbox{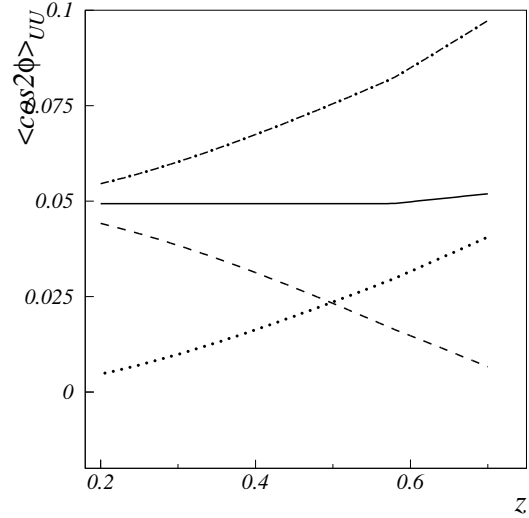}} 
\caption[*]{\label{cos2} The $z$-dependence of the $\cos2\phi$ asymmetry. The 
full and dotted curves correspond to the $T$-even and $T$-odd terms of
asymmetry, respectively. The dot-dashed and dashed curves are the sum and
the difference of those terms, respectively.}
\end{figure}

With spin 1/2 partons one gets~\cite{rav73}: 
\begin{equation}
\frac{W^{00}}{W^{++}}=\frac{4(m^2+\langle k^2_{\perp} \rangle \pm 
\Delta^2 }{Q^2} \qquad \Rightarrow \quad R,
\label{R1}
\end{equation}
\begin{equation}
\frac{W^{+-}}{W^{++}}=\frac{2(\langle k^2_{\perp x}\rangle - 
\langle k^2_{\perp y}\rangle \pm 
\Delta^2 )}{Q^2} \qquad \Rightarrow \quad \langle \cos2\phi \rangle,
\label{R2}
\end{equation}
\begin{equation}
Re\frac{W^{+0}}{W^{++}}=\frac{\sqrt{2}(\langle k_{\perp x} \rangle \pm 
\Delta )}{Q} \qquad \Rightarrow \quad \langle \cos\phi \rangle.
\label{R3}
\end{equation}     
The result (\ref{R1}) corresponds to the ratio $R$ in Eq.(\ref{FB}). 
The predictions that the $\cos\phi$- and  $\cos2\phi$-dependences in 
the cross section behave like $Q^{-1}$ and $Q^{-2}$ should, in practice,  
be easier to test than the ratio in Eq.(\ref{R1}) or Eq.(\ref{FB}). 

The $\langle \cos2\phi \rangle$ from ordinary sub-sub-leading 
$T$-even and leading double $T$-odd (up to a sign) effects 
to order $1/Q^2$ can be written in the form
\begin{equation}
{\langle \cos2\phi \rangle}_{UU} 
=\frac{2\frac{\langle k^2_{\perp }\rangle}{Q^2} (1-y) f_1(x)D_1(z) 
\pm 8 (1-y) h^{\perp (1)}_1(x) H^{\perp (1)}_1(z)}{ \bigg [ 1+{(1-y)}^2 + 
2\frac{\langle k^2_{\perp }\rangle}{Q^2} (1-y) \bigg ] f_1(x) D_1(z)}.
\label{C2PHI}
\end{equation}   

The $z$-dependences of this asymmetry are shown in Fig.~\ref{cos2}. The 
full and dotted curves correspond to the $T$-even and $T$-odd terms in the
asymmetry, respectively. The dot-dashed and dashed curves are the sum and
the difference of those terms, respectively. From Fig.\ref{cos2} one can 
see that the double 
$T$-odd asymmetry behaves like $z^2$, while the $T$-even 
asymmetry is flat in the whole range of $z$.  
Therefore, aside from the
competing $T$-even $\cos2\phi$ effect,  the experimental 
observation of a strong $z$-dependence (especially at high $z$ region) 
would indicate the presence of $T$-odd structures in {\it unpolarized}  
SIDIS implying that novel transversity properties of the  nucleon 
can be accessed  without involving spin polarization.     

\section{Conclusion}
The interdependence of intrinsic transverse
quark momentum and angular momentum conservation are intimately
connected with studies of transversity. 
This was demonstrated previously  
from analyses of the tensor charge in the context
of the axial-vector dominance approach to exclusive meson 
photo-production~\cite{gamb_gold}.  We have analyzed the $T$-odd 
transversity properties in hadrons that emerged in SIDIS. 
We have considered the $\cos2\phi$ asymmetry that appears in unpolarized 
$ep$ 
scattering and have pointed out how its measurements may indicate the 
presence of the asymmetric distributions of transversely polarized quarks 
inside an unpolarized nucleon. Furthermore, we have predicted
a sizeable Sivers asymmetry at HERMES energies. 

Beyond these model calculations it is clear that final state 
interactions can account for  SSAs.  In addition, it has been shown that 
other mechanisms, ranging from  initial
state interactions to the non-trivial phases  of light-cone wave 
functions~\cite{bhs} can account for SSAs.
These various mechanisms can be understood in the
context of gauge fixing as it impacts the gauge link operator
in the transverse momentum quark distribution functions~\cite{ji,ji2}.
Thus using rescattering as a mechanism to generate 
$T$-odd distribution functions 
opens a new window into the theory and phenomenology of transversity in 
hard processes.

\end{document}